\documentclass[prb,twocolumn,showpacs,preprintnumbers,amsmath,amssymb]{revtex4}

\usepackage{graphicx}
\usepackage{dcolumn}
\usepackage{bm}

\begin{document}
\title{Energetics and atomic mechanisms of dislocation nucleation in
strained epitaxial layers}

\author{O. Trushin$^1$}
\address{Institute of Microelectronics and Informatics,
Academy of Sciences of Russia, Yaroslavl 150007, Russia}
\author{E. Granato}
\address{Laborat\'orio Associado de Sensores e Materiais, \\
Instituto Nacional de Pesquisas Espaciais, 12201--970 S\~ao Jos\'e
dos Campos, SP Brasil}
\author{S.C. Ying}
\address{$^2$Department of Physics, Brown University,
Providence, RI 02912 USA }
\author{P. Salo and T. Ala-Nissila$^2$}
\address{$^1$Helsinki Institute of Physics and Laboratory of Physics,
Helsinki University of Technology, FIN--02015 HUT, Espoo, Finland}

\draft

\date{July , 2003}

\begin{abstract}

We study numerically the energetics and atomic mechanisms of
misfit dislocation nucleation and stress relaxation in a
two-dimensional atomistic model of strained epitaxial layers on a
substrate with lattice misfit.  Relaxation processes from coherent
to incoherent states for different transition paths are studied
using interatomic potentials of Lennard-Jones type and a
systematic saddle point and transition path search method. The
method is based on a combination of repulsive potential
minimization and the Nudged Elastic Band method. For a final state
with a single misfit dislocation, the minimum energy path and the
corresponding activation barrier are obtained for different
misfits and interatomic potentials. We find that the energy
barrier decreases strongly with  misfit. In contrast to continuous
elastic theory, a strong tensile-compressive asymmetry is
observed. This asymmetry can be understood as manifestation of
asymmetry between repulsive and attractive branches of pair
potential and it is found to depend sensitively on the form of the
potential.

\end{abstract}


\maketitle

\section{Introduction}

Emergence of misfit dislocations in heteroepitaxial systems is a
long-standing problem in the field of thin film growth
\cite{bean,vdm,matthews,politi00,gil90,dod86,tay87,nan91,much01,gky,tsao}.
Improving the physical properties of semiconductor
heterostructures requires controlling the atomistic processes
responsible for generation of defects. Thus, understanding the
atomistic mechanisms of defect nucleation is crucially important
for further progress in the field of heterostructure growth and
structural control of nanostructures. In addition, misfit
dislocations represent an important problem in fundamental
science. While a lot of information about the nature of
dislocations has been obtained within the traditional continuum
elastic theory, not near as much is known about the details of the
underlying atomistic mechanisms through which dislocation
nucleation occurs.

Energy-balance arguments for the competition between strain energy
build-up and strain relief due to dislocation nucleation in
mismatched epitaxial films lead to the concept of an equilibrium
critical thickness. This is defined as the thickness at which the
energy of the epitaxial state equals that of a state containing a
single misfit dislocation. It has been argued that dislocations
should appear in the film when the thickness exceeds this critical
value \cite{bean,vdm,matthews}. The predicted critical value from
this consideration, however, both from continuous elastic models
\cite{vdm} and from models incorporating layer discreteness
\cite{gky}, is much smaller than the observed experimental value
for the breakdown of the epitaxial state. This suggests that the
defect-free (coherent) state above the equilibrium critical
thickness is metastable \cite{tsao} and the rate of dislocation
generation is controlled by kinetic considerations instead.

The idea of strain relaxation as an activated process is supported
by experimental results for the temperature dependence of the
critical thickness \cite{tsao,luth,zou96} and it is the
fundamental assumption in kinetic semi-empirical models
\cite{hou91}. Physically, the lowest energy barrier for the
nucleation of dislocations should correspond to a transition path
that initiates from the free surface (with or without defects).
Such processes have been considered in  a number of studies using
continuum models \cite{spencer,cullis,grilhe}. However, it has
been pointed out that surface steps and surface roughness that are
not considered in the continuum models could play an important
role for dislocation nucleation \cite{dong98,tersoff94,brochard}.
Thus, atomistic studies are important for a detailed understanding
and determination of the possible mechanisms for defect nucleation
in epitaxial films. Although the importance of kinetic factors in
real experiments have already been emphasized \cite{tsao} and also
investigated in numerical simulations of atomistic models of the
growth process \cite{dong98}, direct determination of the
transition path and corresponding energy barrier for misfit
dislocation nucleation from an epitaxial film has been much less
explored, and they often require assumptions on the particular
structure of the intermediate configuration \cite{ich95}.

The actual stress relaxation processes starting from the epitaxial
coherent state can occur along many different transition paths.
The path with the lowest activation energy barrier at the saddle
point corresponds to the true nucleation barrier for the
generation of a misfit dislocation. For correct determination of
this barrier, it is important to investigate different minimum
energy paths (MEPs) \cite{neb}, from the metastable coherent state
to the incoherent state, without assuming a priori any particular
form of the intermediate configurations. We have recently carried
out such a task which systematically explore the MEPs in the phase
space of the system \cite{tru02a, tru02b} based on a combination
of the Repulsive Bias Potential  (RBP) \cite{tru03} and the Nudged
Elastic Band (NEB) methods \cite{neb}. In the previous work
\cite{tru02a}, we considered the case of a relatively large misfit
of $f=\pm 8$ \%. We showed that there is indeed a nonzero energy
barrier for defect nucleation. Most importantly, however, we
showed that both the mechanisms for the initiation of a misfit
dislocation and the activation barrier exhibit a strong {\it
tensile-compressive asymmetry} which is sensitive to the range of
the interaction potential. A tensile-compressive asymmetry has
also been found previously \cite{dong98,ich95} in other contexts.

In this work, we present a detailed systematic study of defect
nucleation for the same 2D Lennard-Jones system as in Ref.
\onlinecite{tru02a}. We consider strains in the range $f=\pm
(4-8)$ \%, and intermolecular potentials with different ranges.

\section{Model}

\begin{figure}
\includegraphics[bb= 0cm  0cm  7cm   4cm, width= 8cm]{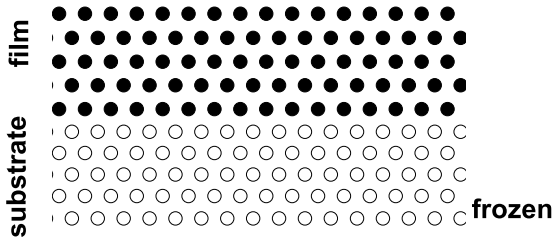}
\caption{A 2D model of the epitaxial film and substrate showing
the particle configurations in the coherent state. The two layers
at bottom are held fixed while all other are free to move. Filled
circles represent the the epitaxial film and open circles the
substrate.  }
\label{model}
\end{figure}

We consider a 2D model of the epitaxial film and substrate where
the atomic layers are confined to a plane as illustrated in Fig.
1. Interactions between atoms in the system were modelled by a
generalized Lennard-Jones (LJ) pair potential \cite{zhen83} that
is modified to ensure that the potential and its first derivative
go to zero at a predetermined cut-off distance $r_c$:
\begin{eqnarray}
&&U(r)=  V(r) , \qquad   r \leq r_0;  \cr &&U(r)=V(r) \left[ 3
\left( \frac {r_c-r}{r_c-r_0} \right) ^2 - 2 \left( \frac
{r_c-r}{r_c-r_0} \right) ^3 \right] , \ r > r_0, \cr
&& \label{LJ}
\end{eqnarray}
where
\begin{equation}
V(r)=  \varepsilon \left[ \frac m{n-m} \left( \frac {r_0}r \right)
^n - \frac n{n-m} \left( \frac{r_0}r \right) ^m \right],
\end{equation}
$r$ is the interatomic distance, $\varepsilon$ the dissociation
energy and $r_0$ the equilibrium distance between the atoms. This
potential for $m=12$ and $n=6$ is the same that has been used by
Dong {\it et al.} \cite{dong98} in a recent simulation study. The
equilibrium interatomic distance $r_0$ was set to a different
value $r_{\rm ss}$, $r_{\rm ff}$ and $r_{\rm fs}$ for the
substrate, film and film substrate interaction respectively. The
parameter $r_{\rm ff}$ was varied to give a a misfit between
lattice parameters as
\begin{equation}
f=(r_{\rm ff}-r_{\rm ss})/ r_{\rm ss}.
\end{equation}
For the film-substrate interaction, we set the equilibrium
distance $r_{\rm fs}$ as the average of the film and substrate
lattice constants, {\it i.e.} $r_{\rm fs}=(r_{\rm ff}+r_{\rm
ss})/2 $. A positive mismatch $f>0$ corresponds to compressive
strain and negative to tensile strain when the film is coherent
with the substrate. Calculations were performed with periodic
boundary conditions in the direction parallel to the
film-substrate interface. For large systems, free boundary
conditions gave qualitative similar results. In the calculations,
the two bottom layers of the five-layer substrate were held fixed
to simulate a semi-infinite substrate while all other layers were
free to move. Typically, in our calculations each layer contained
50 or more atoms. The central portion of the initial epitaxial
film and substrate are shown in Fig.(\ref{model}).

In the previous work \cite{tru02a} it was found that some features
of dislocation nucleation are sensitive to the detailed form of
the atomic potentials used. The results presented here are from
systematic calculations for different values of cut-off distances
for the $5-8$ potential ($m=8$, $n=5$). The advantage of this
potential over the conventional $6-12$ LJ potential is that it is
intrinsically longer ranged. Thus, by imposing different cutoff
radius $r_c$, one can study the influence of the range of the
potential on the nucleation of misfit dislocations. The other
difference with respect to the $6-12$ potential is a softer
repulsive core. This will lead to a weaker anharmonicity and less
asymmetry between the tensile and compressive strain situations.

\section{Method}

The standard way of generating transition paths through Molecular
Dynamics (MD) simulations \cite{ras2000} does not work well in
cases where the probability for rare activated events is small.
There are now numerous methods which have been constructed to
solve this fundamental problem. The MD technique itself has been
augmented by various acceleration \cite{Vot02} and sampling
schemes \cite{Del98,Che99}. In addition, there is a class of
methods that do not evaluate the dynamics directly  but instead
focus on a systematic search of transition paths and related
saddle points for many-particle systems
\cite{Bar96,Hen99,Mun99,relax}.

We have recently introduced \cite{tru03} a particularly simple but
efficient method called the Repulsive Bias Potential (RBP) method
for transition path searching. In the RBP method, the potential
energy of the system is augmented with a fixed, repulsive bias
potential to make the initial configuration unstable, but to keep
the other nearby minima unaffected:
\begin{equation}
U_{\rm tot}(\vec r,\vec r_0)=U(\vec r)+A\exp\{-[(\vec r-\vec
r_{0})/\alpha]^2\}.
\end{equation}
Here $U(\vec r)$ is the original potential energy surface of the
system, which has been modified by an exponentially decaying,
spherically symmetric potential of strength $A$ and range $\alpha$
which is centered at $\vec r_{0}$. When $A$ and $\alpha$ have been
chosen appropriately, forces computed from Eq. (1) can be used to
displace the system from its initial state located at $\vec r_{0}$
to escape to a nearby minimum. This is done by applying total
energy minimization to $U_{\rm tot}$.

With the RBP method implemented, the procedure of determining the
transition path comprises several stages. First, the initial
epitaxial state is prepared by minimizing the total energy of the
system using MD cooling. In the MD cooling method, the energy is
gradually minimized by setting the velocities ${\bf v}=0$ whenever
${\bf v}$ and the force ${\bf f}$ on a particle satisfy the
condition $ {\bf v } \cdot {\bf f} < 0 $. Positions and velocities
of the particles are obtained from numerical integration of the
equations of motion using the standard leap-frog algorithm.
Following this, the RBP is applied and the system is slightly
displaced from the initial state (randomly or in a selective way
to escape from harmonic basin) and then total energy minimization
is applied to find a new minimum energy state.

It is important to note that the RBP method can generate many
different final states depending on both the initial displacements
and the exact form of the repulsive bias introduced. By making the
repulsive bias sufficiently localized around the initial potential
minimum, the final state energy depends only on the true potential
of the system and not on the fictitious repulsive bias. In this
work, we only consider the final configurations corresponding to
the presence of a single misfit dislocation. Rather than trying
random initial displacements, some knowledge of the dislocation
generation mechanism is useful for expediting the process.

We also find that the proper choice of initial displacements
depends on the sign of the misfit. In the case of compressive
strain, to get an ideal single dislocation located in the center
of our sample, the optimal  initial displacement corresponds to
moving one atom in the middle of the first layer of the film from
the film-substrate interface upwards by a small distance ($0.04
r_{\rm ss}$ ). In case of tensile strain, the corresponding
optimal initial displacement is a small displacement ($0.04 r_{\rm
ss}$ ) downwards for an atom located in the middle of the second
layer in the film from the film-substrate interface layer.

While the repulsive bias potential minimization can be used to
generate the final state configuration containing a misfit
dislocation, it does not yield the precise minimum energy path and
the lowest activation barrier value for getting to this final
state configuration. For this purpose, we use the Nudged Elastic
Band (NEB) method \cite{neb}. This is an efficient method for
finding the minimum energy path (MEP) , given the knowledge of
both initial and final states. The MEP is found by constructing an
initial set of configurations (images) of the system between the
initial and final states. This set is then allowed to relax to the
true set representing the MEP.

An initial guess for the images in the NEB is usually obtained by
interpolating the particle configurations between the final and
initial states. For the present application, however, we find that
this often leads to numerical instabilities due to the strong hard
core repulsion of the LJ potentials and fail to converge to the
true MEP. To circumvent this problem, we use the set of
configurations generated in moving to the final state in the
presence of the repulsive bias as the initial input in the NEB.
This leads to fast convergence in the NEB method without the
instabilities encountered in the linear interpolation scheme.

\section{Results}

For epitaxial films above the equilibrium critical thickness, the
relaxed state with a nonzero density of misfit dislocations which
partially relieves the strain energy in the film is expected to
have a lower energy. However, if this configuration is separated
from the coherent state by a finite energy barrier $\Delta E$, the
film will remain coherent unless defects are nucleated allowing to
overcome this energy barrier. This barrier could be finite even
when the relaxed state  has already a lower energy than the
epitaxial state. Thus the experimentally observed critical
thickness  can be much larger than the equilibrium value depending
on the kinetics of defect nucleation. Our preliminary results
\cite{tru02a,tru02b} showed a large variety of relaxation
processes, including single dislocation nucleation, multiple
dislocations, dislocations with different core structures, and
dislocations nucleating on different depth in the film, which can
be characterized by their different activation energies and
energies of the final incoherent states. In this work, we focus on
the nucleation and MEP leading to a final state containing only a
single misfit dislocation with core located near the
film-substrate interface. To simplify the discussions, we will
present in this section only the results for the $5-8$ potential
with a cutoff radius of $r_c=1.5 r_{\rm ss}$, and lateral size
$L=50$, corresponding to $50$ atoms per layer. These results allow
us to arrive at a simple physical picture for the nucleation
process of the misfit dislocation. The results with different
parameters for the intermolecular potential and different size of
the system are qualitatively similar.  They will be presented in a
later section.

\subsection{Mechanisms of relaxation}

\begin{figure}
\includegraphics[bb= 0cm  0cm  9cm   8cm, width= 8cm]{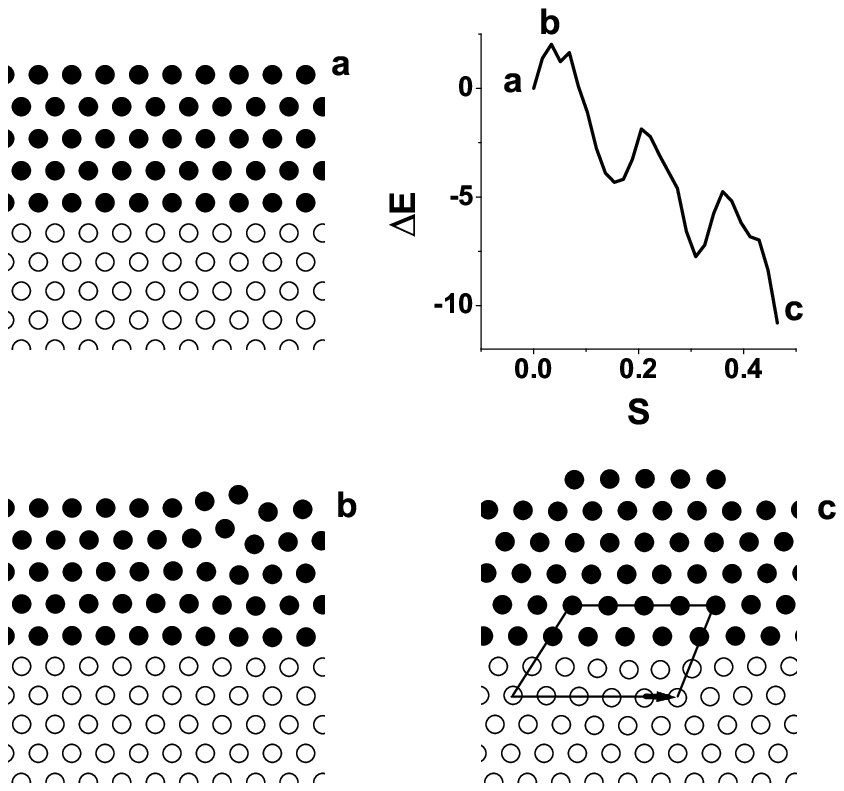}
\caption{ Minimal Energy path for compressive strain $f=+8\%$ as a
plot of energy barrier $\Delta E$ vs reaction coordinate S.
Snapshots configurations (a), (b) and (c) correspond to the labels
in the energy profile (top right). Closed line in (c) is the
Burgers circuit around the dislocation core. The energy barrier is
in units of interatomic potential strength $\epsilon$ and the
reaction coordinate $S$ is in units of equilibrium distance
$r_{\rm ss}$. } \label{pathc}
\end{figure}

\begin{figure}
\includegraphics[bb= 0cm  0cm  9cm   8cm, width= 8cm]{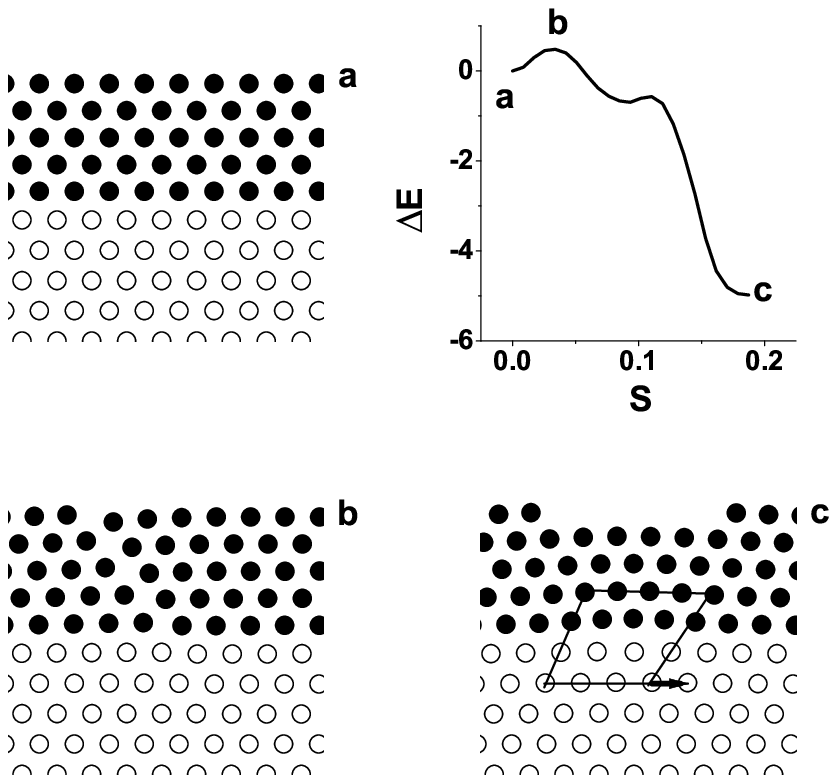}
\caption{ Minimum Energy  path for tensile strain $f=-8\%$ as a
plot of energy barrier $\Delta E$ vs reaction coordinate S.
Snapshots configurations (a), (b) and (c) correspond to the labels
in the energy profile (top right). Closed line in (c) is the
Burgers circuit around the dislocation core. The energy barrier is
in units of interatomic strength $\epsilon$ and the reaction
coordinate $S$ in units of equilibrium distance $r_{\rm ss}$. }
\label{patht}
\end{figure}

Relaxation of strain with dislocation nucleation is a complex
process involving motion of many particles inside the system. The
transition from coherent to dislocated state considered in this
paper is analogous to strain relaxation in a real heteroepitaxial
sample under annealing conditions. Experiments show that heating
is a essential prerequisite for such relaxation to occur
\cite{tsao,zou96}. This fact shows that nucleation of dislocation
represents typical activated process with a nonzero activation
barrier. Our calculations with NEB confirm this conclusion
\cite{tru02a,tru02b}. For both the compressive and tensile strain
cases, we find the presence of a finite activation barrier $\Delta
E$ along the MEP leading from the initial epitaxial state to the
final state with a single misfit dislocation in the film substrate
interface. To allow for comparison of different cases and
extraction of the basic physics involved, we introduce the
definition of the reaction coordinate $S$. This is defined as the
accumulated displacement of the system along the MEP in the
multidimensional configuration space. Mathematically, the reaction
path coordinate for a given configuration (image) along the MEP is
defined as

\begin{equation}
S_M =  \sum_{m=1}^M\sqrt{\sum_{i=1}^N(r_i^m - r_i^{m-1} )^2 /N}
\label{reactionc}
\end{equation}
Here M is the label for the configuration (image ) under
consideration, and i is the index for the different particles in
the system (i=1 to N). In Fig.(\ref{pathc}) and Fig.(\ref{patht}),
we show typical snapshots of configurations along the
corresponding MEP for compressive and tensile strain cases
respectively. In all cases the initial state was an epitaxial film
with a coherent interface and the final state contained a single
dislocation with its core located in the interface layer. The
final state is characterized by the presence of an adatom island
on the surface of the film in the case of compressive strain and a
vacancy island in the tensile case. The number of adatoms (or
vacancies) in the island exactly corresponds to the number of
layers in the film. Such form of the final state is determined by
the geometry of the misfit dislocation, as the
 one extra atom is added or removed from (or inside) each layer to
relax the strain.

An important property of the NEB method is that it usually
converges to the MEP nearest to the initial trial trajectory. Thus
by changing the initial input  path, we were able to investigate
several different mechanisms of relaxation \cite{tru02a,tru02b}.
These mechanisms differ from each other mainly by the level of
collectiveness in the displacement of the particles from the
coherent state position. For each given set of parameters, we
identify the lowest activation barrier. The particular kind of
mechanism leading to the lowest activation barrier depends on the
parameters of model (misfit, cut-off radius of potential {\it
etc.}). We find that for all the systems that we have studied, the
mechanisms leading to the lowest activation barrier belong to one
of the two categories described below.

\begin{figure}
\includegraphics[bb= 0cm  0cm  9cm   8cm, width= 8cm]{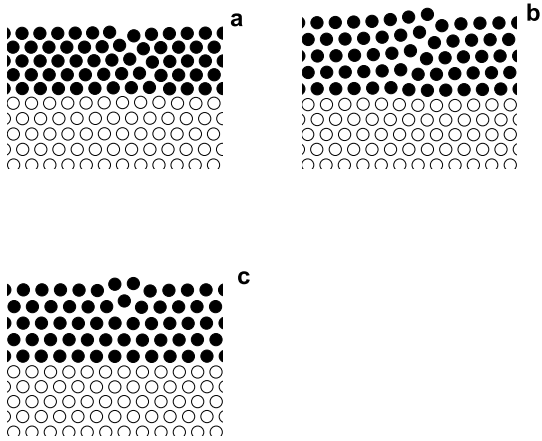}
\caption{ Saddle point configurations for different mechanisms of
stress relaxation   (a) glide mechanism for tensile strain (b)
glide mechanism for compressive strain (c)  climb mechanism for
compressive strain.  Filled circles represent the the epitaxial
film and open circles the substrate.}
 \label{mech}
\end{figure}

The first mechanism describing the transition from the initial
coherent state to the final state with a misfit dislocation at the
film substrate interface  corresponds to a successive sliding
along the edges of a triangle. The saddle point configurations
corresponding to this mechanism for the tensile and compressive
strain cases are shown in Fig. (\ref{mech}a) and Fig.
(\ref{mech}b) respectively.  We see that in this case the
displacements of the atoms have a collective behavior, with two
edges of a triangle successively sliding up or down (one by one).
Eventually, an adatom island or a vacancy island is created on the
surface of the film. The highest saddle point can correspond
either to the sliding of the first or the second edge. We refer to
this as the glide mechanism since the motion of the dislocation
after it is nucleated follows the path referred in the literature
as dislocation glide \cite{politi00}. For the tensile strained
film, the glide mechanism always yield the lowest activation
barrier. While for the compressively strained film, the mechanism
leading to the the lowest activation barrier depends actually on
the magnitude of the misfit. For small misfit ( $f \leqslant 8
\%)$, the glide mechanism is again the one leading to the lowest
activation barrier. This is drastically different from the climb
mechanism reported earlier \cite{tru02a} for a misfit of $8\%$ in
a compressively strained film.

The second mechanism correspond to successive relaxation of
layers. This is the preferred mechanism for a compressively
strained film with large misfits ($f\geqslant 8 \%$). The saddle
point configuration
 corresponding to this mechanism for the compressive strain
of 8\%  misfit  is shown in Fig.(\ref{mech}c).
 In this case,  the core of the
dislocation  first  appears at either the second or the third
layer of the film and then successively moves down from layer to
layer to the film-substrate interface. The displacement of the
particles have a very localized character in this kind of
mechanism. We refer to this  as the climb mechanism since the
motion of the dislocation after it is first nucleated in this case
corresponds to what is  known in the literature as
 dislocation climb \cite{politi00}. For intermediate values of compressive
strain, the situation is more complicated, as the two mechanisms
are competitive in energy costs. The actual MEP  in this case is
better described by a mixture of the climb and glide mechanisms.


\subsection{Activation energy of dislocation nucleation}

The most important characteristic of a particular relaxation
process through nucleation of a misfit dislocation is its
activation energy $\Delta E$. The activation barrier is calculated
as the difference between the total energy of the initial state
and that of the saddle point configuration. As can be seen in Fig.
(\ref{pathc}), corresponding to the compressive strain case, there
may exist many saddle points along a given MEP . The activation
barrier is determined by the highest energy saddle point.  The
results for $\Delta E$ {\it vs.} the number of layers in the film
are presented in Fig.(\ref{barrier1}).

For the tensile strain case, we find that the process leading to
the nucleation of misfit dislocation and subsequent motion along
the MEP is always through the glide mechanism. The activation
barrier decreases with increasing magnitude of misfit. Also, at
large misfits, the activation barrier decreases significantly as
the film thickness increases, leading to an essentially negligible
activation barrier. This was verified directly through independent
MD simulation at finite temperatures where the misfit dislocation
is easily generated spontaneously.

For the compressive strain case, except at $4\%$ misfit and small
thickness (less than six layers), the barriers are higher than the
corresponding tensile strain case with the same magnitude of
misfit. Again, there is a strong decrease in $\Delta E$ with
increasing magnitude of misfit. In contrast to the tensile strain
case, the activation barrier tends to level off with increasing
film thickness. The other striking difference from the tensile
strain case is that the mechanism corresponding to the movement
along the MEP in this case can either be the glide mechanism as in
the tensile strain case, or the qualitatively totally different
climb mechanism involving layer by layer distortion as discussed
in the last section. This new climb mechanism occurs for large
misfits ($ f \geqslant 8\%$).

\section{Simple Physical Picture for the Nucleation Process}

As shown in the last section, the mechanism leading to the
nucleation of a misfit dislocation starting from the epitaxial
coherent state and the subsequent motion along the MEP to the
final state is fairly complicated, and depends sensitively on the
sign and magnitude of the misfit (tensile or compressive strain),
 and thickness of the film.  With this rich set
of data, it is important to have some simple qualitative
understanding of the results.

\begin{figure}
\includegraphics[bb= 0cm  0cm  6cm   6.5cm, width= 8cm]{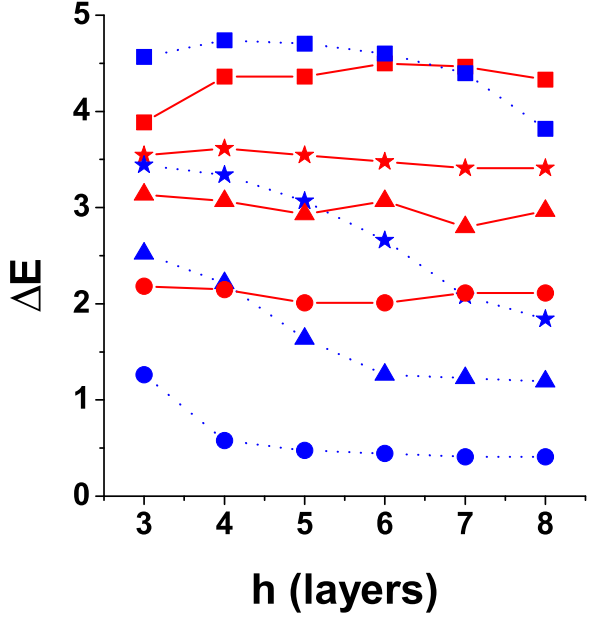}
\caption{ Energy barrier $\Delta E$ (in units of $\epsilon$) as a
function of film thickness (number of layers) for different misfit
values. Squares symbols correspond to $f=\pm 4 \%$, stars to
$f=\pm 5\%$, triangles to $f=\pm 6\%$, and circles to $f=\pm 8\%$.
Solid and dotted lines correspond to compressive $f > 0$ and
tensile $f < 0$ strains, respectively. } \label{barrier1}
\end{figure}

\begin{figure}
\includegraphics[bb= 0cm  0cm  6cm   6.5cm, width= 8cm]{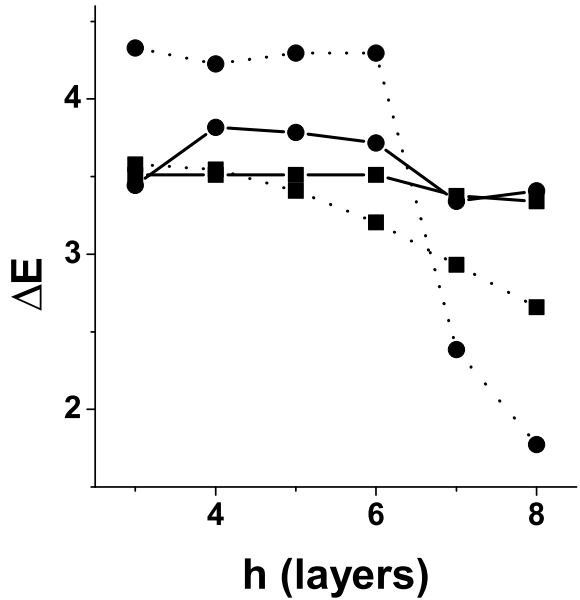}
\caption{Energy barrier $\Delta E$ (in units of $\epsilon$) as a
function of film thickness (number of layers) at  misfit $5\%$,
for the $5-8$ (squares) potential and $6-12$ (circles) potential
(cut off $1.5 r_{\rm ss}$). Solid and dotted lines correspond to
compressive $f > 0$ and tensile $f < 0$ strains, respectively.
Here the system size is $L=20$.} \label{LJb}
\end{figure}

First of all, it is easy to understand the origin of the
difference between the tensile and compressive strain cases. In a
harmonic elasticity theory, the activation barrier would depend
only on the magnitude and not the sign of the strain. The
tensile-compressive asymmetry thus originates from the strong
anharmonicity of the interaction potential, particularly in the
steeply rising repulsive core. This is confirmed by our results
shown in Fig.(\ref{barrier1}) showing that the difference of
$\Delta E$ for the tensile and compressive cases grows
monotonically as the misfit increases in magnitude. This is also
confirmed in our similar studies using the conventional $6-12$ LJ
potential as shown in Fig.(\ref{LJb}). Since the $6-12$ potential
is considerably steeper in the core region, the anharmonicity is
stronger and the resulting tensile-compressive asymmetry is even
more pronounced.

\begin{figure}
\includegraphics[bb= 0cm  0cm  6.5cm   11cm, width= 8cm]{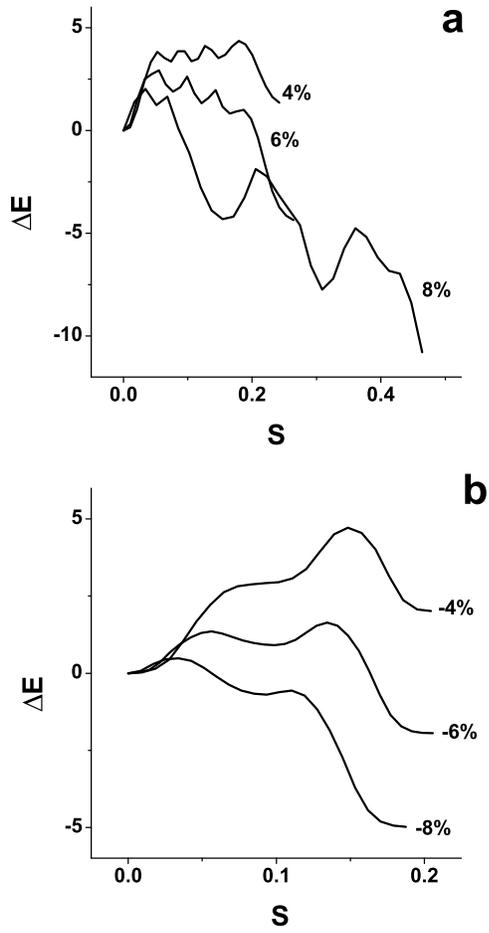}
\caption{Energy profile of the minimum energy path for (a)
compressive  and (b) tensile  strain and for different misfits.
Energy in units of $\epsilon$ and $S$ in units of equilibrium
distance $r_{\rm ss}$.}
 \label{position}
\end{figure}

The other general trend is the strong decrease of the activation
barrier with increasing misfit. This is true for both the tensile
and compressive cases (Fig. \ref{barrier1}). It remains true even
when the mechanism leading to the nucleation has changed character
from a glide nature to a climb nature as in the case of large
compressive strain. In our previous work \cite{tru02a}, we have
analyzed the contribution to the activation barrier from the
intralayer and interlayer bond distributions at the saddle point.
Here we will introduce the same physical arguments in terms of the
conceptually simpler quantity of reaction coordinate defined
earlier in Eq. (\ref{reactionc}). Let $S$ represent the
dimensionless reaction coordinate along the MEP leading from the
initial coherent state through the saddle point to the final state
containing the misfit dislocation. For the initial stages of small
displacement with $S \ll 1$, the simplest leading representation
of the MEP can be expressed in the form
\begin{equation}
E = \frac{a}{2} S^2 - \frac{b}{3} S^3. \label{reaction}
\end{equation}
%
%
In the equation above, the first term gives the energy rise
towards the saddle point from the initial displacements from the
coherent state necessary to nucleate the dislocation. It
originates mainly from the stressing of the interlayer bonds which
are fully relaxed in the initial coherent epitaxial state. Because
of this initial relaxation, there is relatively little dependence
of the coefficient $a$ on the misfit. The second term represents
the release of the intralayer strain energy from the displacements
of the atoms. Clearly, the coefficient $b$ is strongly dependent
on the magnitude of the misfit. Whether it is tensile or
compressive, the higher the magnitude of the strain, the larger is
the lowering of the strain energy. Hence the coefficient $b$
should be a monotonically increasing function of the magnitude of
the misfit. It follows simply from Eq. (\ref{reaction}) that the
activation barrier $\Delta E$ is given by the expression
\begin{equation}
\Delta E = \frac{1}{6} \frac{a^3}{b^2}.
\end{equation}
Thus, the activation barrier always decreases with increasing
magnitude of the strain, whatever the actual initial strain
release mechanism and nature of the saddle point configuration.
Furthermore, the expression in Eq. (\ref{reaction}) predicts that
the saddle point should occur at the reaction coordinate $S_0=a/b$
which again  decreases monotonically as the misfit magnitude
increases. This is supported by our results as shown in
Fig.(\ref{position}).

In general, the initial cost of energy in creating the distortion
for the dislocation in the glide mechanism  is lower for a tensile
than compressive strain. This is due to the fact that for the
compressive strained film, the initial distortion required for
creating the dislocation core always involve breaking of bonds to
lower the coordination number. On the other hand,  for the tensile
strained film, no breaking of bonds is necessary in the glide
mechanism for the nucleation of the dislocation. Thus the glide
mechanism is always preferred for the tensile strained film.  For
the large compressive strain, the energy cost  involved in
nucleating a dislocation core is comparable for  the glide and
climb mechanism, and the two processes are competitive.

The dependence of the activation barrier on the film thickness is
more complicated and is rather different for the tensile and
compressive strains. For the large compressive strain case where
the MEP corresponds to the climb mechanism, the behavior is fairly
easy to understand as the saddle point involves a rather localized
dislocation in the surface layers, so obviously the activation
barrier would have a very weak dependence on the film thickness as
observed in our numerical study. For the glide mechanism, both the
initial rise in energy and the release of the strain energy
leading to the saddle point configuration are dependent on the
film thickness and according to Eq. (\ref{reaction}), it is hard
to predict any universal dependence of the activation barrier on
the film thickness. Indeed, both a levelling off (for compressive
strain) and a strong decrease in the activation barrier as a
function of the film thickness have been observed.

\section{ Size and Potential Dependence}

The results presented in the previous sections are all for a $5-8$
short ranged L-J potential with a cutoff set at $1.5 r_{\rm ss}$.
The size of the system was set at $L=50$ particles per layer. We
have also performed similar calculations for different set of
parameters in the potential as well as for different sizes to
investigate the size and potential dependence of our results. We
find  that the results with different interatomic potentials and
sizes of the system are qualitatively similar, although differing
in details. We present some of these results in this section.

\begin{figure}
\includegraphics[bb= 0cm  0cm  6cm   6.5cm, width= 8cm]{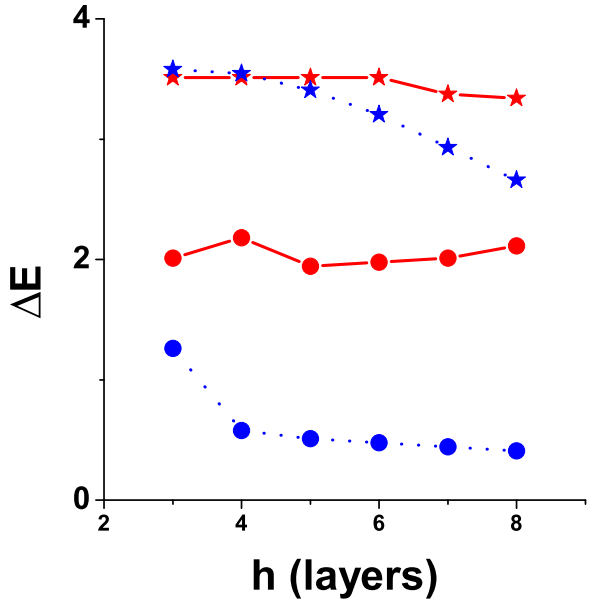}
\caption{ Energy barrier $\Delta E$ (in units of $\epsilon$)  as a
function of film thickness (number of layers) for smaller sample
size ($20$ atoms per layer) and different misfit values for the
$5-8$ potential:  $f=\pm 5\%$ ( stars), and  $f=\pm 8\%$
(circles). Solid and dotted lines correspond to compressive $f >
0$ and tensile $f < 0$ strains, respectively. } \label{delta}
\end{figure}

In Fig. (\ref{delta}), the activation energy barrier is plotted
against the film thickness for a system size of $L=20$ and a short
ranged potential as in the previous sections for two values of the
magnitudes of misfit at $|f|=5\%$ and $8\%$. The results are very
similar to that presented in Fig. (\ref{barrier1}). The only
limitation for the smaller sample size is that one cannot
accurately study the cases of smaller misfit as the addition or
removal of a single atom from a layer would overshoot the strain
release mechanism.

\begin{figure}
\includegraphics[bb= 0cm  0cm  6cm   6.5cm, width= 8cm]{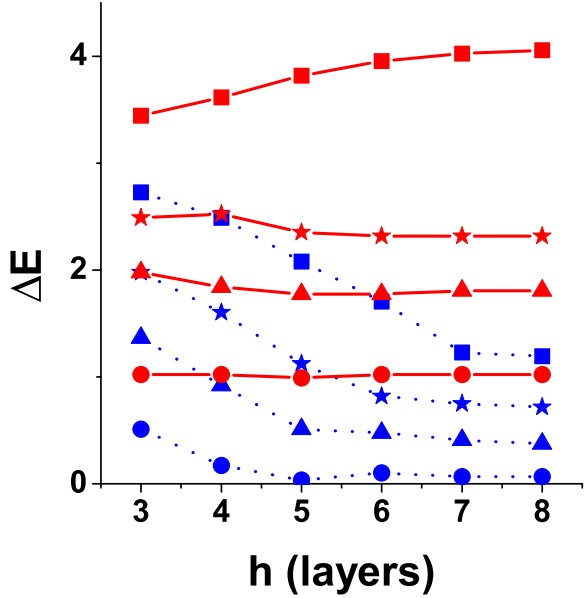}
\caption{
 Energy barrier $\Delta E$  (in units of $\epsilon$) as
a function of film thickness (number of layers) for different
misfit values for long ranged $5-8$ potential (cut off $2.1 r_{\rm
ss}$): $f=\pm 4 \%$(squares), $f=\pm 5\%$ (stars),  $f=\pm 6\%$
(triangles), and $f=\pm 8\%$ (circles). Solid and dotted lines
correspond to compressive $f> 0$ and tensile $f < 0$ strains,
respectively. } \label{barrier2}
\end{figure}

In Fig. (\ref{barrier2}), we show the results of activation energy
barrier vs film thickness for system size $L=50$ and a $5-8$ LJ
potential as before but this time with a longer range with cutoff
set at $2.1 r_{\rm ss}$. Again, the results are qualitatively
similar to that presented in Fig. (\ref{barrier1}). The tensile
and compressive asymmetry is stronger for this longer ranged
potential, particularly at the smaller misfit values. This could
be also related to the stronger size effects for the longer ranged
potential.

\section{Conclusions}

We have developed a general scheme of identifying minimal energy
paths for spontaneous generation of misfit dislocation in an
epitaxial film and studied the energetics and atomic mechanisms of
stress relaxation using a two-dimensional model. This approach
requires no a priori assumptions about the nature of the
transition path or the final states. A nonzero activation barrier
for dislocation nucleation is found in the minimum energy path
from coherent to incoherent state. We find that the energy barrier
decreases strongly with  misfit. The nucleation mechanism from a
flat surface depends crucially on whether we start from a tensile
or compressive initial state of the film. This asymmetry
originates from the anharmonicity of the interaction potentials
which leads to qualitatively different transition paths for the
two types of strains. The present method can also be extended to
three-dimensional models with more realistic interaction
potentials. Preliminary calculations for a three-dimensional
Lennard-Jones system and the Pd/Cu and Cu/Pd systems \cite{unp}
with the Embedded Atom Model potentials \cite{eam} confirms the
effectiveness of the method in three dimensions.

\section{Acknowledgements}

This work has been supported in part by Funda\c c\~ao de Amparo
\`a Pesquisa do Estado de S\~ao Paulo - FAPESP (grant no.
03/00541-0) (E.G.), the Academy of Finland through its Center of
Excellence program (T.AL and P. S.) and by a  NSF-CNPq grant (E.G.
and S.C.Y.).

\newpage


\end{document}